\documentclass[aip,jcp,preprint,floatfix]{revtex4-1}
\pdfoutput=1 
\usepackage{comment} 
\usepackage{subfigure}
\usepackage{float}
\usepackage{epsfig}
\usepackage{booktabs}
\usepackage{amsmath}
\usepackage{amssymb}
\usepackage{color}
\usepackage{setspace}
\usepackage{comment}
\usepackage{subfig}
\usepackage{geometry}
\usepackage{indentfirst}
\usepackage{bbold}
\usepackage{graphicx}

\begin{document}
\title{How Electronic Dynamics with Pauli Exclusion Produces Fermi-Dirac Statistics}

\author{Triet S. Nguyen}
\affiliation{Department of Chemistry and Biochemistry, University of Notre Dame, Notre Dame, IN 46556}
\author{Ravindra Nanguneri}
\affiliation{Department of Chemistry and Biochemistry, University of Notre Dame, Notre Dame, IN 46556}
 {
\author{John Parkhill}
\affiliation{Department of Chemistry and Biochemistry, University of Notre Dame, Notre Dame, IN 46556}
\email{jparkhil@nd.edu}

\date{\today}

\begin{abstract}
    It is important that any dynamics method approaches the correct population distribution at long times. In this paper, we derive a one-body reduced density matrix dynamics for electrons in energetic contact with a bath. We obtain a remarkable equation of motion which shows that in order to reach equilibrium properly, rates of electron transitions depend on the density matrix. Even though the bath drives the electrons towards a Boltzmann distribution, hole blocking factors in our equation of motion cause the electronic populations to relax to a Fermi-Dirac distribution. These factors are an old concept, but we show how they can be derived with a combination of time-dependent perturbation theory and the extended normal ordering of Mukherjee and Kutzelnigg. The resulting non-equilibrium kinetic equations generalize the usual Redfield theory to many-electron systems, while ensuring that the orbital occupations remain between zero and one. In numerical applications of our equations, we show that relaxation rates of molecules are not constant because of the blocking effect. Other applications to model atomic chains are also presented which highlight the importance of treating both dephasing and relaxation. Finally we show how the bath localizes the electron density matrix.\end{abstract}

\maketitle

\section{Introduction}
\indent     A perfect theory of electronic dynamics must relax towards the correct equilibrium distribution of population at long times. However, satisfying this condition is not trivial. The two most common approaches for simulating non-equilibrium electronic dynamics in chemistry, Ehrenfest method and the surface hopping approach, do not satisfy this condition exactly. In the case of Ehrenfest dynamics, the violation of detailed balance is severe\cite{Ehrenfest} and the dynamics can only be taken seriously at short times. In the case of surface hopping, the violation is so small it is hard to identify\cite{shdetbal}. Both these methods are usually formulated in a basis of distinguishable (Boltzmannian) electronic states. Because the basis must be limited in practice, predictions in the Boltzmannian picture depend on a choice of adiabatic or quasi-diabatic states. States with fractional orbital occupations are also extremely expensive and difficult to treat in a basis of collective electronic states. To address these problems, we are pursuing relaxation dynamics based on the reduced electron density matrices without wavefunctions.\\
\indent   In the one-electron picture, non-equilibrium Green's function theories\cite{PhysRevC.58.1594,PhysRevB.46.4757,Semkat,neqGF} and Boltzmann transport equations\cite{Boltzz,erpa} do satisfy detailed balance. Master equations which exchange both electrons and energy between a system and a bath of two-level Fermions have also been recently derived and approach a Fermi-Dirac (FD) distribution\cite{PhysRevA.90.022123,2014arXiv1410.0304S}. However, for spectroscopy the most important bath is undoubtedly the bosonic vibrations of the atoms\cite{PhysRevB.89.085202}. To our knowledge, it has never been explicitly shown how a bosonic bath drives electrons to their correct equilibrium as presented in this paper. Here we show how a Pauli-blocking effect slows down relaxation, and dephasing of electrons. The derivation that we use is general and can be applied to electron correlation as well. We also use our new expressions to show that the rate of relaxation for a particle-hole excitation is not constant because of the Pauli-blocking effect. \\
\indent     Ordinary system-bath perturbation theory with a bath of bosons does not tend towards FD populations. The equations do not consider exchange statistics so they approach a canonical equilibrium instead. The goal of this paper is to show how the derivation of ordinary system-bath perturbation theory naturally leads to a FD equilibrium if expectation values are taken correctly with respect to a mixed-state, Fermionic vacuum. The resulting kinetic equations resemble Redfield theory\cite{redfield} but also depend on the hole density. The key to this paper is that the Fermionic expectation value is taken using the extended normal ordering technique of Mukherjee and Kutzelnigg (MK) \cite{Kutzelnigg:1997aa} which doesn't require a wavefunction and only depends on reduced density matrices (RDMs). The exact equation of motion (EOM) for many-Fermion RDMs, the BBGKY \cite{PhysRevB.85.235121}, is also non-linear, and we show that this non-linearity is also a requirement to obey Fermionic detailed balance. \\
\indent     The application of Mukherjee and Kutzelnigg's powerful technique to finite temperature quantum mechanics is natural. They hinted at this application in their classic paper, and Hirata and He recently invoked the technique to correct perturbation theory in the thermodynamic limit\cite{Hirata:KohnLutt}. Unfortunately, the generalized normal ordering technique is challenging to derive, understand, and apply. We must refer the reader to the original papers for details. A related generalized Wick's theorem\cite{orig_Wick} concept was recently developed for Green's function-based theories\cite{PhysRevB.85.115119}. \\
\indent    For linear response away from a determinant, the blocking effect is irrelevant\cite{Tempel:2011aa} and so the literature is basically correct. A paper using these new formulas in a useful all-electron dynamics code is an immediate follow up. Electron transport methods based on the density matrix will also benefit from the main results of this paper\cite{Zelovich:2014kq,Chen:2014fj,Subotnik:2009aa,ThossKinetics}. Electronic statistics can be easily enforced as long as kinetic rates depend on the density matrix. 

\section{Detailed balance}
\indent   In the non-interacting case, the population of an electronic state at long times should tend to the Fermi-Dirac (FD) distribution:
\begin{align}
\bar{n}_i = \frac{1}{1+e^{\beta(\epsilon_i-\mu)}}
\label{eq:db}
\end{align}
where $\beta = (k_bT)^{-1}$, $\epsilon_i$ is the energy of a one-electron state, and $\mu$ the chemical potential satisfying $\sum_i \bar{n}_i=N$, the total number of electrons. The bar indicates that this is an equilibrium quantity, where the number of particles exchanged between any two states are equal so the corresponding populations do not change (the condition of detailed balance). \\
\indent     It is known that a kinetic equation can be easily modified by 'Pauli-blocking' factors $(1-n_i)$, to reach detailed balance with the FD distribution. Consider a set of non-negative Markovian transition rates between states $K_{ij}$ satisfying $\sum_j K_{ij}=1$, and canonical detailed balance $e^{\beta(\epsilon_j-\epsilon_i)}=K_{ij}/K_{ji}$. The blocking factors close any channel which would disobey exclusion. Inserting FD $\bar{n}_i's$ into a kinetic equation of this form:
\begin{align}
    \frac{d}{dt} n_i = \sum_j K_{ij} (1-n_i) n_j = \sum_j K_{ij} \eta_i n_j
\label{eq:dbferm}
\end{align}
one sees the number of particles exchanged between any two states are equal, meaning the FD distribution is stationary. We have introduced $\eta_i$, the diagonal of the one-hole density matrix. Equation (\ref{eq:dbferm}) is the form a kinetic equation must take to satisfy Fermionic detailed balance. Exclusion factors are an old concept attributed to Landau\cite{landau}, as an \textit{ad hoc} correction. They are also important for dephasing rates\cite{imry}. In the remainder of this paper, we derive an expression which reduces to Eq. (\ref{eq:dbferm}) with perturbation theory. 

\section{Fermionic Master Equation}
\indent We begin our derivation with the following Hamiltonian: 
\begin{align}
    \hat{H} = \hat{H}_{el} + \hat{H}_b+ \hat{V} = \sum_{i} \epsilon_i a^i a_i + \sum_{\alpha}\omega_\alpha b^\alpha b_\alpha +\sum_{i,j,\nu} V_{i,j}^\nu a^ia_j \hat{B}_\nu
\label{eq:sbham}
\end{align}
Where $\hat{B}_\nu$ is an coupling-weighted sum of boson operators attached to $\nu$, defined below. To show that FD distribution is the exact equilibrium, the electrons must be non-interacting, but the correction we derive applies in the interacting case as well. We use a notation for second quantization which follows Ref. \cite{Kutzelnigg:1997aa} and summation over repeated indices is implied. 
In the open-systems approach, the Hamiltonian is put into this form after diagonalizing the electronic part from some localized basis $|i\rangle =\sum_j \hat{U}_{i\nu} |\nu\rangle$ with electronic coupling. In the original basis, which is the atomic basis in our work, the electronic part of the system-bath coupling is usually taken to be diagonal:
\begin{align}
\hat{V} = \sum_\alpha g_{\alpha,\nu}\omega_\alpha(b^{\alpha,\nu}+b_{\alpha,\nu})|\nu \rangle \langle \nu| = \hat{B}_\nu |\nu \rangle \langle \nu|
\end{align}
but leads to off-diagonal coupling: $V^\nu_{i,j} = \sum_\nu {U}_{i,\nu}{U}^\dagger_{j,\nu}$ in the system energy eigenbasis used to propagate the dynamics. We discuss how Eq. (\ref{eq:sbham}) is prepared atomistically in the Results section. The equilibrium Boson expectation values taken in this work are denoted $\langle \rangle_\text{eq}$, otherwise $\langle \rangle $ denotes a Fermionic expectation value. No restriction is made on the nature of the Fermionic state which is used to perform the expectation value. In particular, the formulas are valid for a general mixed Fermionic state which is completely defined by its RDMs and \emph{no wavefunction is invoked or implied}. We also define a projection operator $\mathcal{P}$, which replaces Bosonic operators with their expectation values at equilibrium leaving the Fermion operators untouched, e.g. $\mathcal{P} a^{a}a_{b}b^\dagger b=a^{a}a_b \langle b^\dagger b \rangle_\text{eq}$, and its complement operator $\mathcal{Q} = \hat{I} - \mathcal{P}$.  \\
\indent 	The one-electron reduced density matrix (1-RDM) is defined as $\gamma^a_b = \langle a^{a}a_b \rangle= \langle a^a_b \rangle$. We assume that an initial 1-RDM we would like to propagate is known, perhaps on the basis of a stationary electronic structure calculation. In the notation of this work which follows that of Ref. \cite{Kutzelnigg:1997aa}, $\gamma$ is not an operator. Rather it is the expectation value of a one-particle operator. 1-RDMs are not equivalent to the tight-binding (TB) density matrices assumed in a master equation. However, Fermionic \emph{operators} do satisfy Heisenberg's EOM like TB density matrices. The steps used to produce master equations for TB operators generically apply to one-body Fermionic operators as well. \textbf{Thus until fermionic expectation values are taken, our derivation is essentially the same as the textbook derivations of Redfield theory that can be found in Breuer\cite{breuer_theory_2002} or Nitzan\cite{nitzan2006chemical}}. Our final expressions also reduce to the commonly used second-order TB theory in the limit that Fermi-blocking factors are removed.\\
\indent     Working in the interaction picture with respect to the first two terms of $\hat{H}$, where $\mathcal{L}(t') = -i[\hat{V}(t'),\cdot]$, second-order time convolutionless perturbation theory\cite{parkhill_correlated-polaron_2012,Shibata:1977tg, breuer_theory_2002} produces this EOM for the tensor product of $a^a_b$ with an equilibrium boson state: 
\begin{align}
\frac{d}{dt} \mathcal{P}a^a_b(t) =  \left( \mathcal{PL}(t)\mathcal{P} + \int_{0}^t dt'  \mathcal{PL}(t)\mathcal{Q}\mathcal{L}(t')\mathcal{P} \right ) a^a_b(t) + \mathcal{I}(t)
\label{eq:tcl2}
\end{align}
Taking the Fermionic expectation value of Eq. \ref{eq:tcl2} will eventually produce an EOM for $\gamma$, because $\frac{\partial}{\partial t}\gamma^a_b = \frac{\partial}{\partial t} \langle a^{a}_b \rangle $, but the equilibrium Boson expectation values in $\mathcal{P}$ must also be taken\footnote{Also note that an equation of motion for $\gamma$ alone is not enough to exactly propagate an \emph{interacting} many-Fermion state which would require equations for higher RDMs. We assume non-interacting electrons, but our equations would be compatible with a time-dependent Hartree-Fock approximation.}. The first term on the right of Eq. (\ref{eq:tcl2}) is zero because it is sandwiched by $\mathcal{P}$ and the expectation value of odd-numbered boson operators are zero at equilibrium. The last term is an inhomogeneity which accounts for any initial correlations between electrons and bosons. Our previous work explored this term in detail \cite{parkhill_correlated-polaron_2012}. The effects on observables in that work were shown to be small relative to the Markov approximation, and other papers document how inhomogeneous effects decay \cite{PhysRevA.86.062114,breuer_theory_2002}. Because we are interested in the long-time detailed balance limit, we are well motivated to neglect the inhomogeneity\cite{PhysRevA.88.052107}. This leaves only the second term, reflecting the time-dependent correlation between the system and the bath, which we denote as $\frac{d}{dt} (a^{a}_b)^{(2)}$:
\begin{align}
 \frac{d}{dt} (a^{a}_b)^{(2)}(t)  = (\frac{i}{\hbar})^2 \big\langle \int_{0}^t dt'\big [[a^{a}_b(t),V^\nu_{c,d}a^c_d(t')\hat{B}_\nu(t')],V^\nu_{e,f} a^{e}_f(t) \hat{B}_\nu(t)] \big\rangle_\text{eq}
 \label{eq:term2}
\end{align}
At this point we move back to the Schrodinger picture. We collect the time-dependent factors in each term ($e^{i\omega_{cd}t'}, e^{i\omega_{ef}t}$ where $\hbar \omega_{ab}=\epsilon_a-\epsilon_b$) and interaction strength scalars, $V^\nu_{a,b}$ together with the boson expectation value to focus on the Fermionic expectation value. An EOM for $\gamma$ is obtained by taking the Fermonic expectation value of both sides of this equation. On the left hand side, $\frac{d}{dt} e^{i\omega_{ab}t} a^a_b = i\omega_{ab}e^{i\omega_{ab}t}a^a_b + e^{i\omega_{ab}t} \frac{d}{dt} a^a_b$. Dividing the common factor and taking the expectation value, $i\omega_{ab} \langle a^a_b\rangle = i\omega_{ab}\gamma^a_b$ produces the first term in equation 10.\\
\indent Ordinarily, in order to take an expectation value of products of operators like on the right hand side of Eq. \ref{eq:term2}, one needs a description of an electronic state as excitations relative to a determinantal wavefunction. Instead we use MK's extended normal ordering which advantageously only depends on RDMs, but requires too much algebra to publish in detail. We wrote a Mathematica program to take the expectation value which is available from us upon request. This double commutator also occurs in MK's paper\cite{Kutzelnigg:1997aa} as equation (48). To obtain our expression from theirs, we first delete terms in which contain operators (whose expectation values are therefore zero), and also make a simple substitution ($\delta^a_b = \eta^a_b+\gamma^a_b$, with $\eta$ the hole density matrix and $\gamma$ the particle density matrix) so that $\delta^a_b \gamma^c_d - \delta^c_d \gamma^a_b = \eta^a_b\gamma^c_d - \eta^c_d\gamma^a_b$, leaving four terms:
\begin{align}
 \big\langle \big[[a^a_b,a^c_d],a^e_f\big] \big\rangle = \delta^c_b(\gamma^a_f\eta^e_d-\eta^a_f\gamma^e_d)+\delta^a_d(\gamma^e_b\eta^c_f-\eta^e_b\gamma^c_f)
\end{align}
This expression has two interesting consequences. First, it shows that the expectation value of this double commutator does not depend on the higher-order RDMs. Even if the reference had strong, two-particle correlations, or it was a double-excitation that \emph{does not} affect this term\footnote{See MK's paper for details.}. This result supports the possibility of an accurate dissipative time-dependent Hartree-Fock equation since the dissipation of the one-body density is not coupled to the two-body density \emph{through the bath}.\footnote{The two-body density does couple to the one-body density via the Coulomb interaction in an atomistic Hamiltonian.} Secondly, each factor in this EOM has a naturally derived Pauli-blocking factor.\\
\indent     The remaining steps in our derivation are equivalent to the ordinary derivation of a Redfield equation. If $\eta$ is replaced with the identity matrix (blocking is shut off) even the result is equivalent to an ordinary Redfield equation, and so readers can consult with one of the excellent textbooks\cite{nitzan2006chemical}. The first assumption is a Markov approximation ($\lim_{t\rightarrow \infty}$), in which the bath remains at Boson equilibrium. This places a constraint on the boson correlation function: 
\begin{align}
\int_0^\infty \big\langle \hat{B}_\nu(t)\hat{B}_\nu(0) \big\rangle_{eq} e^{i\omega_{ij} t} dt= \Gamma_\nu(\omega_{ij}) = e^{\beta\hbar\omega_{ij}}\Gamma_\nu(\omega_{ji})
\end{align}
Which leads to the canonical detailed balance relation in the Redfield equation, but for this equation furnishes canonical K's like Eq. (\ref{eq:dbferm}). Putting the result of the time integration together with the Fermionic expectation value for each term we obtain:
\begin{align}
 \frac{d}{dt} \big\langle (a^{a}_b)^{(2)} \big\rangle(t) = (\frac{i}{\hbar})^2 \Big\{ \Gamma_\nu(\omega_{cf})V^\nu_{ae}V^\nu_{fc}\gamma^c_b\eta^e_f+
\big(\Gamma_\nu(\omega_{df})V^\nu_{be}V^\nu_{fd}\big)^\dagger \gamma^a_d\eta^e_f \\ \notag - \Gamma_\nu(\omega_{ca})V^\nu_{de}V^\nu_{ac}\gamma^c_d\eta^e_b -
\big(\Gamma_\nu(\omega_{db})V^\nu_{ce}V^\nu_{bd}\big)^\dagger \gamma^c_d\eta^a_e \Big\}
\label{eq:etcl}
\end{align}
We note an interesting duality between the electron and the hole in this expression. The third and fourth terms are particle relaxing, but their blocking factors resemble hole-dephasing terms (they couple to only one index of $\eta$), whereas the first and second are particle dephasing, but resemble hole-relaxing terms.\\
\indent A secular approximation must be made to reach an equation which preserves trace, by deleting contributions which couple the diagonal and off-diagonal elements of the $\gamma$. Besides the desired trace preservation, the secular approximation can be motivated with a rotating wave argument, which is identical to the Redfield case described in several previous publications\cite{PhysRevB.88.174514,muka}. Unit-less Kronecker deltas enforce this decoupling $\text{S}_{a,b,c,d}=\delta_{ab}\delta_{cd}+\delta_{bc}\delta_{ad}(1-\delta_{ab}\delta_{cd})$. The resulting electronic EOM is the key result of our paper:
\begin{align}
\dot{\gamma_b^a} = \frac{-i}{\hbar}\omega_{ab}\gamma_b^a + \frac{-1}{\hbar^2} \Big\{ \text{S}_{a,f,f,c}\Gamma_\nu(\omega_{cf})V^\nu_{ae}V^\nu_{fc}\gamma^c_b\eta^e_f+\\ \notag
\text{S}_{b,f,f,d}\big(\Gamma_\nu(\omega_{df})V^\nu_{be}V^\nu_{fd}\big)^\dagger \gamma^a_d\eta^e_f- \text{S}_{d,b,a,c}\Gamma_\nu(\omega_{ca})V^\nu_{de}V^\nu_{ac}\gamma^c_d\eta^e_b-\\ \notag
\text{S}_{c,a,b,d}\big(\Gamma_\nu(\omega_{db})V^\nu_{ce}V^\nu_{bd}\big)^\dagger \gamma^c_d\eta^a_e \Big\}
\label{eq:fulleom}
\end{align}
which is similar to the form of a secular Redfield equation but with blocking factors. We propagate this non-linear equation using a straightforwards Runge-Kutta-Fehlberg adaptive integrator. $\eta$ is not propagated separately, it is simply always set equal to it's definition in terms of $\gamma$. The equation rigorously preserves the trace of the 1-RDM and equilibrates to FD distribution. In the case that the density starts without coherence, $\eta$ is diagonal and the equation takes the form of Eq. (\ref{eq:dbferm}). We will now discuss the importance of these blocking factors in the evolution of populations and coherences using calculations on model systems.
\section{Results}
\indent     First, we will demonstrate that propagation of Eq. (\ref{eq:etcl}) reproduces FD statistics. To produce a model with the form of Eq. (\ref{eq:sbham}), we consider a Huckel model of atomic sites. Each site has an energy $E_\nu=1.0$ au and is coupled to its nearest neighbors with a coupling strength $\hat{V}^e_{\nu,\nu}$. The energy of each site fluctuates due to the system-bath coupling, and the spectrum of these fluctuations is determined by a spectral density for this site $J_\nu(\omega)=\sum_\alpha \hbar^2\omega_\alpha^2 g_{\nu,\alpha}^2 \delta(\omega-\omega_\alpha)$ of the Drude-Lorentz form. We have previously shown a method to obtain these spectral densities atomistically from limited amounts of molecular dynamics, but in this work we assume a strong, high-frequency bath\footnote{Four Drude-Lorentz functions are chosen with widths, amplitudes and central frequencies of the following form (au): \{(0.0001, 0.00001, 0.0001),  (0.001, 0.0017, 0.0017), (0.013, 0.023, 0.027), (0.01, 0.017, 0.017)\}} to test positivity and rapidly achieve relaxation. After assigning a spectral density and coupling in the real-space $\nu$ basis, the electronic Hamiltonian is diagonalized to recover a Hamiltonian of the form (\ref{eq:sbham}). Some of the dynamics is performed at a very high temperature so that the exact reproduction of FD-statistics is made clear. However, blocking has important effects even at room temperature; we show some dynamics results at 300K to this effect. In all cases, we half-fill the chain with electrons, and the chemical potential of the system is constant as per Eq. (\ref{eq:sbham}). \\
\begin{figure}
\centering
\includegraphics[width=0.55\textwidth]{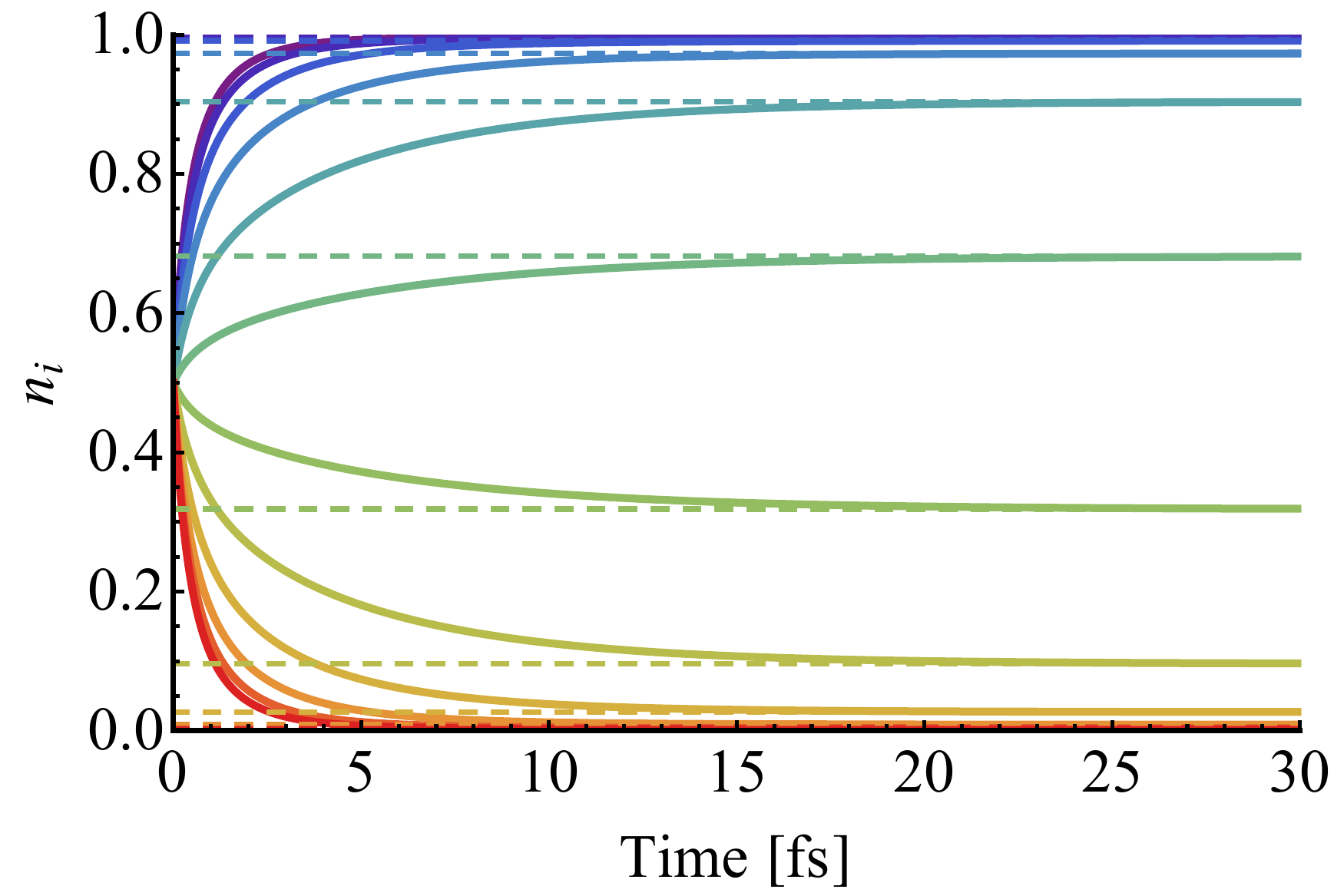}
\caption{A linear tight binding chain with 12 sites, half-filled with 6 electrons, initially at infinite temperature, in which populations of all states are equal, approaches a FD distribution at 500 K in the long time limit ($\hat{V}^e_{\nu,\nu} = 0.005$ au). Solid lines denote the energy eigenstate populations as a function of time; the dashed lines are the exact asymptotes, the Fermi-Dirac population of each eigenstate given the energy and chemical potential.}
\label{fig:fermi}
\end{figure}
\indent     The electronic dynamics resulting from Eq. (\ref{eq:etcl}) preserves the trace of the electronic density matrix and produces a FD distribution at long times. Experiments with several different coherent and incoherent initial conditions and a very strong bath all equilibrate to a FD distribution (Fig. \ref{fig:fermi}). Solutions of the ordinary secular Redfield equation are exponential with time constants determined by the eigenvalues of $K_{ij}$; however, our trajectories with Fermi blocking do not take the same form. There are direct implications for the kinetics of excited state relaxation, and the interpretation of transient absorption spectra which are ordinarily fit to a superposition of decaying exponentials. Because of Fermi-blocking, particle-hole excitations and especially multiple particle-hole excitations do not relax with simple exponential kinetics. We define a ground-state probability based on the overlap of the T=0 K ground state and the density during the dynamics: $P_g(t)=\prod_i (1-|n_i(t)-\Theta(\epsilon_i-\mu)|)$, where $\Theta$ is the step function. Here we simulate an 8-site chain at T=300 K, with otherwise unchanged parameters. We initialize the density in a particle-hole excited state by unfilling HOMO-2 and filling HOMO+2 and monitor $P_g(t)$ to observe the time dependence of the relaxation rates (Fig. \ref{fig:rlxation}). 
\begin{figure}
\includegraphics[width=1\textwidth]{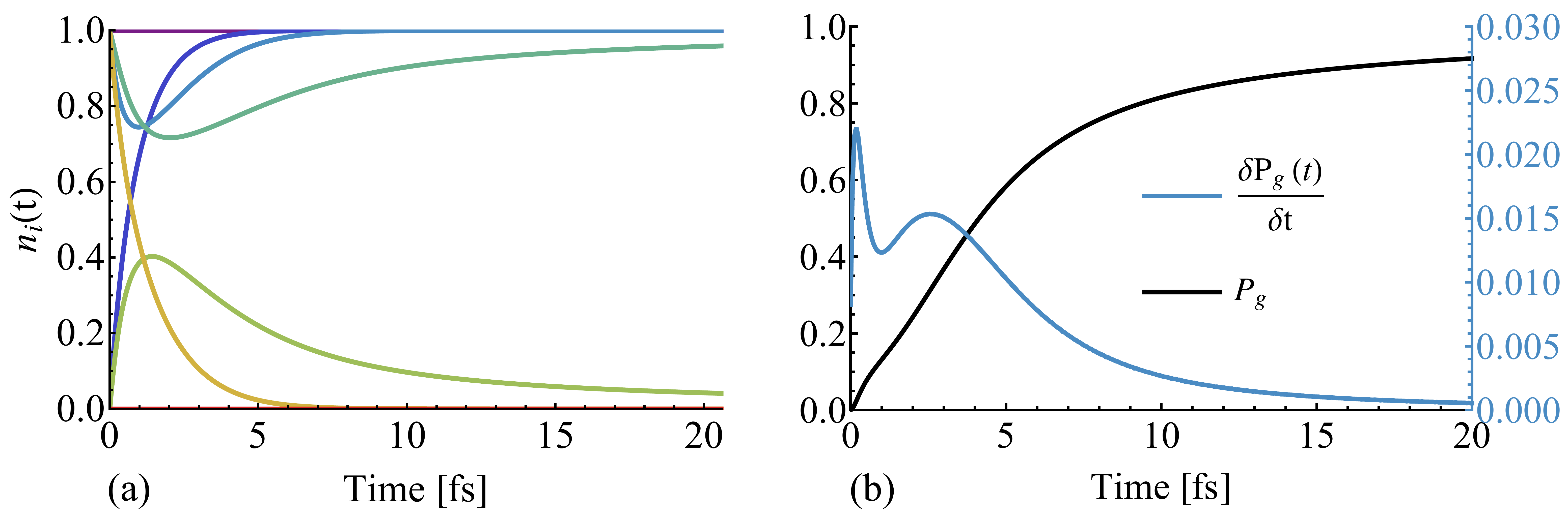}
\caption{An 8-site, 4-electron model is initialized in a particle-hole state (HOMO-2 $\rightarrow$LUMO-2), and allowed to relax at 300 K ($\hat{V}^e_{\nu,\nu} = 0.05$ au). (a) Energy eigenstate populations as functions of time. (b) The rate (right scale) at which the model returns to the many-electron ground state (population on the left scale) is not constant as a function of time due to blocking effects.}
\label{fig:rlxation}
\end{figure}
For single particle-hole states, the rate of relaxation begins rapid as electrons fill in holes (Fig. \ref{fig:rlxation}) and then adopts a slower 'blocked' and essentially exponential relaxation rate dominated by LUMO$\rightarrow$HOMO relaxation at long times. The effective rate only varies by roughly a factor of ten because of the blocking effect. Consequently, it would be difficult to separate from other possible rate fluctuations in ultrafast spectroscopic experiments although we predict that it exists.\\
%
%
\indent Off-diagonal elements of a density matrix are called coherences, and their decay is called dephasing. As has been recently reported by other authors, the dephasing rate of Fermions is considerably different from what ordinary Redfield theory would predict\cite{imry}. In distinguishable models of electronic relaxation, coherence tends to be an ambiguous quantity because it is dependent on the chosen basis. In both the energy eigenbasis, and the position eigenbasis coherence has important and measurable implications. In the former, coherence should be zero in the thermodynamic limit. When they are nonzero, energy-basis coherences oscillate. Our kinetic model does completely decay these coherences and also shows why an atomistic and improvable model for their decay is so important. If we initialize electrons on the left of our chain to reach detailed balance, they naturally flow to the right (Fig. \ref{fig:decohpop}). However if only populations are allowed to relax, even though the populations of energy eigenstates rapidly reach a perfect FD distribution, the undecayed coherences lead to unphysical oscillatory currents in the system. \\
\begin{figure}
\centering
\includegraphics[width=0.83\textwidth]{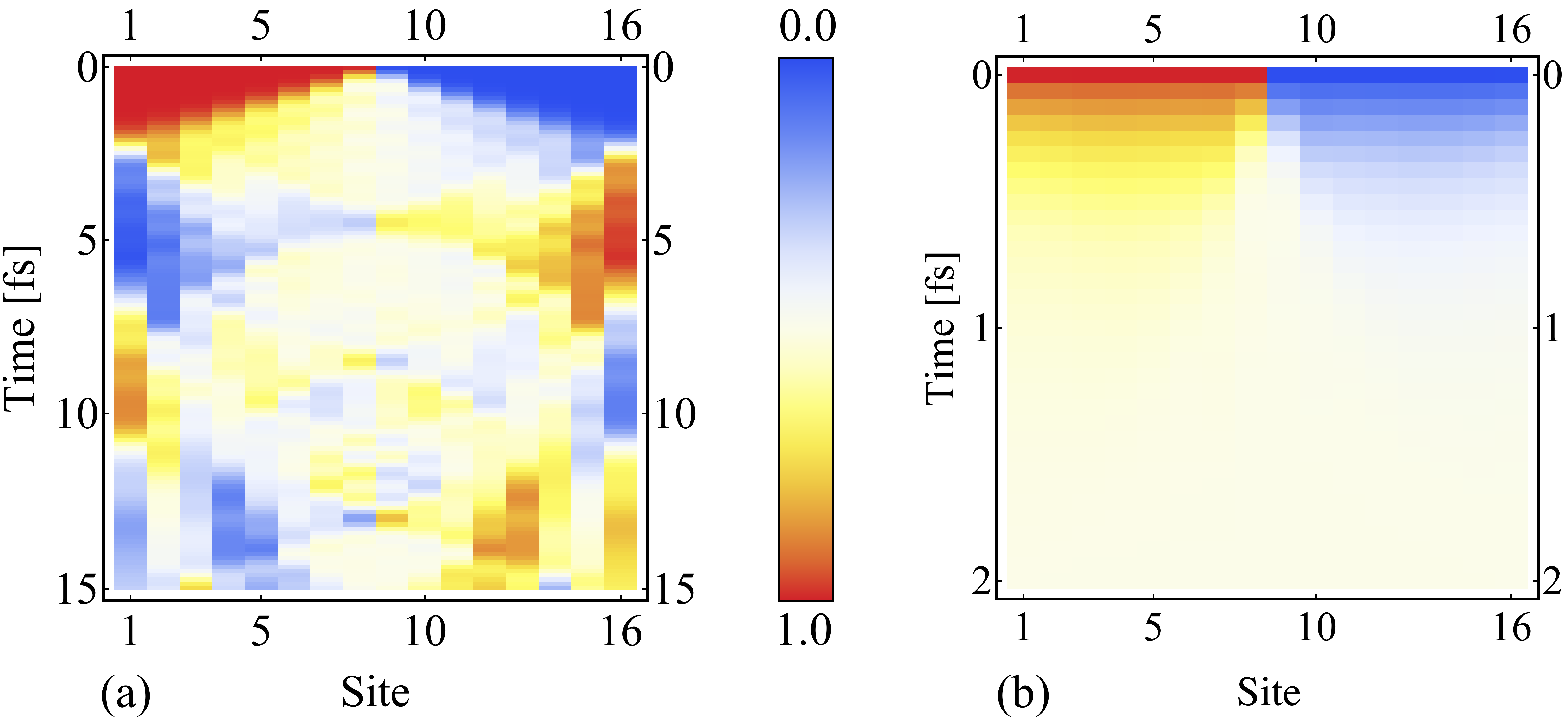}
\caption{The atom-basis populations of two trajectories at 5000 K ($\hat{V}^e_{\nu,\nu} = 0.05$ au) (a) without dephasing and (b) with dephasing, showing the importance of dephasing in electronic dynamics. In both cases, the initial condition is half-filling with all electrons forced to the left of the chain. The populations in the energy eigenbasis of these trajectories are rigorously FD at final time (the bottom of the plot). Without dephasing (a) electron currents do not decay, whereas with all the terms in Eq. (\ref{eq:etcl}) (b) they rapidly decay and all electronic motion is halted by dissipation to the bath after the initial relaxation. Dephasing is as important to accurate electronic dynamics as population kinetics.}
\label{fig:decohpop}
\end{figure}
\indent  In the position eigenbasis, coherences determine the measurable quantum delocalization of particles. Real-space coherence also limits the efficiency of electronic structure theory, since the cost of calculating exchange energy is completely determined by how slowly the real-space density matrix decays away from the diagonal. With our new theory, we can predict how dynamics and finite temperature effects can lead to the long-range decay of the off-diagonal elements of the electronic density matrix (Fig. \ref{fig:localization}). This locality can potentially be exploited to make electronic structure models more efficient. 
\begin{figure}
\centering
\includegraphics[width=0.83\textwidth]{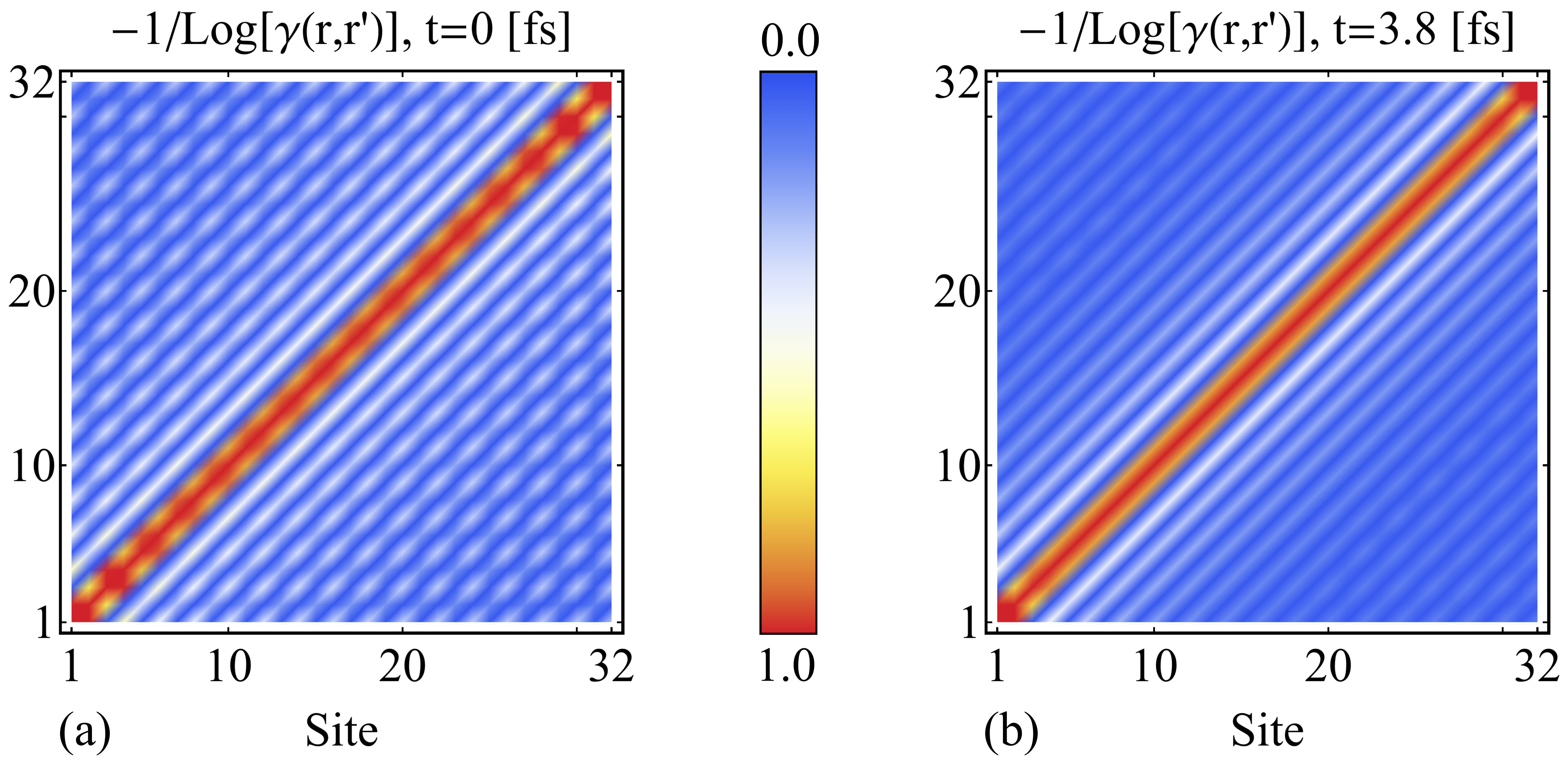}
\caption{A 32-atom chain is initialized into its ground state ($\hat{V}^e_{\nu,\nu} = 0.05$ au), and allowed to warm in contact with a 5000 K bath. $-1/\text{Log}\{\gamma(x,x')\}$ at two different times are plotted at the same scale to make the coherence decay visible.}
\label{fig:localization}
\end{figure}
Although kinetic equation (\ref{eq:dbferm}) is non-linear in the density matrix, in our experience it faithfully maps any density matrix to FD equilibrium at long times, as it should since FD density is a known solution of Eq. (\ref{eq:etcl}). We have not proven that FD is the \emph{only possible solution}, but we strongly suspect it is. 

\section{Discussion and Conclusions}
\indent Fermionic relaxation rates must depend nonlinearly on the density matrix to be correct at long times. Consequently, relaxation rates for Fermionic states are not constant as a function of time, and tend to slow down over the course of non-radiative relaxation. This effect is extremely unclear in a Boltzmannian 'exciton' model. Hole blocking makes rates of electronic relaxation and dephasing slower than the corresponding rates for distinguishable particles.  \\
\indent  These are second-order equations. The rates of transition they predict will become inaccurate in the strong coupling limit, even though we obtain FD with very strong couplings. The main feature of these equations is that they are systematically improvable. One can easily 'Fermi-block' surface hopping rates and perform surface hopping for individual electrons. The results of this paper suggest that with Fermi blocking surface hopping for electrons would, for all intents and purposes, tend to FD. However, that electronic surface hopping would beg for a systematically improvable path to dephasing rates, like its Boltzmann counterpart. \\
\indent     We have found that the MK normal-ordering is an especially useful technique for dynamics, allowing us to work around the impurity of the state. We used it to provide a promising EOM for electrons and simulate non-equilibrium vibronic relaxation. As in our previous work, the perturbation theory could also be used to develop electronic equations of motion that treat electron correlation effects, although that will require an automated version of ENO algebra. This paper opens the door to answering whether interacting equilibrium differs meaningfully from the FD equilibrium in real materials, and how that equilibrium is approached. We are currently extending this work towards higher electronic density matrices. \\
\indent     It is significant that these equations approach Fermionic detailed balance naturally, as opposed to enforcing detailed balance in an \textit{ad hoc} way. In a related vein, sometimes Pauli-blocking factors are taken to be $\bar{\eta}_i$ instead of $\eta_i(t)$, which is correct at equilibrium. However this work suggests that using the equilibrium value is only appropriate in the long time limit. Applications of this work to predict rates of non-radiative relaxation in molecules is an active pursuit in our laboratory. 

\section{Acknowledgements} We thank The University of Notre Dame's College of Science and Department of Chemistry and Biochemistry for generous start-up funding and Honeywell Corporation. T. S. N. gratefully acknowledges the support of the NSF Graduate Research Fellowship under grant 1313583. We acknowledge valuable conversations with Liviu Nicolaescu (Notre Dame), Geoff Thompson (Iowa State), Jarrod McClean, Samuel Blau, and Thomas Markovich (Harvard).

\bibliography{fermi}

\end{document}